\documentclass[12pt]{article}

\usepackage{PRIMEarxiv}

\usepackage[utf8]{inputenc} 
\usepackage[T1]{fontenc}    
\usepackage{hyperref}       
\usepackage{url}            
\usepackage{booktabs}       
\usepackage{amsfonts}       
\usepackage{nicefrac}       
\usepackage{microtype}      
\usepackage{lipsum}
\usepackage{fancyhdr}       
\usepackage{graphicx}       
\usepackage{subcaption} 
\usepackage{amsmath} 
\usepackage{amssymb}
\usepackage[style=numeric,sorting=none,maxnames=1,backend=bibtex]{biblatex}
\graphicspath{{media/}}     
\DeclareUnicodeCharacter{2009}{\,}

\title{A new possible way to detect \\
Axion Anti-quark Nuggets
}

\author{
  I. Lazanu, M. Parvu$^*$ \\
University of Bucharest, Faculty of Physics \\
POBox MG-11 Bucharest-Magurele, Romania\\
  $^*$ Corresponding author:\texttt{ mihaela.parvu@unibuc.ro}
}

\begin{document}
\maketitle

\begin{abstract}
The axion anti-quark nugget (A$\Bar{\mathrm{Q}}$N) model was developed to explain in a natural way the asymmetry between matter and antimatter in Universe. In this hypothesis, a similitude between the dark and the visible components exists. The lack of observability of any type of dark matter up to now, in particular A$\Bar{\mathrm{Q}}$Ns, requires finding new ways of detecting these particles, if they exist. In spite of strong interaction with visible matter, for such objects a very small ratio of cross section to mass is expected and thus huge detector systems are necessary. This paper presents a new idea for the direct detection of the A$\Bar{\mathrm{Q}}$Ns using minerals as natural rock deposits acting as paleo-detectors, where the latent signals of luminescence produced by interactions of A$\Bar{\mathrm{Q}}$Ns are registered and can be identified as an increased and symmetrical deposited dose.
The estimates were made for minerals widely distributed on Earth, for which the TL signal is intense and if the thermal conditions are constant and with low temperatures, the lifetime of the latent signals is kept for geological time scales.
\end{abstract}


\section{Introduction}

A large fraction of the matter in the Universe is not directly observable, its origin is not yet understood, and it is generically referred to as Dark Matter (DM). The existence of DM is established by observations from galactic to cosmological scales, where many phenomena can be better understood if the presence of additional (unobserved) matter is assumed. In order to understand what DM truly is, it is necessary to unambiguously observe its possible interactions with ordinary matter. Regarding the origins and characteristics of DM, a wide variety of particles have been predicted in different theories or phenomenologically, ranging from very massive macroscopic objects like primordial black holes and MACHOs, to cold DM, warm DM, milli-charged particles, self-interacting DM, Weakly Interacting Slim Particles (WISPs), Ultra-light (fuzzy) Dark Matter, or Weakly Interacting Massive Particles (WIMPs).
Other examples are gravitinos, axions, neutralinos, strangelets, Q-balls or sterile neutrinos. For a more complete review see for example Cirelli et al. \cite{cirelli2024dark}. There is a wide variety of experiments searching for axions, QCD-axions, and axion-like particles. These include ADMX \cite{PhysRevLett.120.151301}, CAST \cite{PhysRevLett.112.091302}, ALPS \cite{ortiz2021design}, HAYSTAC \cite{PhysRevLett.118.061302}, MADMAX \cite{lee2020status}, IAXO \cite{vogel2013iaxo}, and CASPEr \cite{Garcon_2017}, each employing unique techniques to probe the existence of these elusive particles.
Notable examples of experiments that are actively searching for WIMPs include the LUX \cite{Akerib_2013} experiment, CDMS \cite{PLBrink_2009} and SuperCDMS \cite{supercdmscollaboration2022effective}, the XENON1T \cite{Aprile_2023} experiment, and the AMS-02 \cite{Jorge_Casaus_2009}.

Predicting the existence of Axion (anti)Quark Nuggets  as exotic constituent particles of dark matter (DM) has a long history, beginning with the pioneering work of Witten \cite{Witten:1984rs}, De Rujula and Glashow \cite{DeRujula:1984axn}, until their detailed definition by Zhitnitsky \cite{Zhitnitsky:2002qa}.  The DM and Baryogenesis are two sides of the same coin; apparently, these two phenomena are uncorrelated. In fact, in the Big Bang model, the existence of axion quark nuggets and axion anti-quark nuggets respectively can restore the symmetry between matter and antimatter, if the total baryon numbers satisfy the following relations:

\begin{equation}
    B_{\text{visible}}:B_{\text{nuggets}}:B_{\text{antinuggets}} \approx 1:2:3
\end{equation}
\begin{equation}
    B_{\text{tot}}=0=B_{\text{nuggets}}-B_{\text{antinuggets}}+B_{\text{visible}}
\end{equation}

Very important, the basic available constraints from Big Bang nucleosynthesis (BBN) and cosmic microwave background (CMB) are not in contradiction with existence of this form of antimatter \cite{SinghSidhu:2020cxw}.

Oaknin and Zhitnitsky \cite{PhysRevD.71.023519} proposed a novel scenario to explain the observed cosmological asymmetry between matter and antimatter which relies on a mechanism of separation of quarks and antiquarks in two coexisting phases at the end of the cosmological QCD phase transition. Some fraction of them are bound into heavy nuggets of quark matter in a colour superconducting phase. Nuggets of both matter and antimatter are formed as a result of the dynamics of the axion domain walls \cite{Zhitnitsky_2003}.
The observed difference between the quantities of nuggets ($\Omega_{\mathrm{N}}$) and antinuggets ($\Omega_{\bar{\mathrm{N}}}$) generated by CP violation unequivocally suggests that the baryon contribution ($\Omega_{\mathrm{B}}$) must be of the same order of magnitude as $\Omega_{\bar{\mathrm{N}}}$ and $\Omega_{\mathrm{N}}$. This is because all these components are proportional to the fundamental dimensional parameter $\Lambda_{\mathrm{QCD}}$, as are all other dimensional parameters in QCD, such as the color superconducting (CS) gap ($\Delta$), critical temperature ($T_c$), and chemical potential ($\mu$). The remaining antibaryons in the early universe plasma then annihilate, leaving only the baryons. The antimatter counterparts of these baryons are bound in the excess of antiquark nuggets and are thus unavailable for fast annihilation. Since all asymmetry effects are of order one, this ultimately leads to similarities among all components, both visible and dark.  The observed ratio of matter to dark matter, $\Omega_{\mathrm{DM}} \approx 5 \cdot \Omega_{\mathrm{B}}$, implies a scenario where the baryon charge concealed in antinuggets exceeds that in nuggets by a factor of approximately $(\Omega_{\bar{\mathrm{N}}} /\Omega_{\mathrm{N}}) \approx 3/2$ at the end of nugget formation \cite{Zhitnitsky_2021}. Recently, Sebastian Baum et al. proposed a systematic direction for investigation of neutrino interactions and searches for exotic particles predicted in the extensions of the Standard Model or only phenomenological, using effects in different minerals and rocks \cite{BAUM2023101245}.

Starting from the reference work by Polymeris et al. \cite{POLYMERIS2006207}, the current article proposes the detection of axion antiquark nuggets - complex systems using the spatial distribution of the deposited doses in various minerals from natural deposits.

\section{The structure of the A$\Bar{\mathrm{Q}}$Ns and their interactions}
Inside the A(anti)QN exists more regions with distinct structures and different length scales; for details see for example \cite{Lawson:2013bya} and \cite{Lazanu_2024}. In accord with Gorham and Rotter \cite{Gorham:2015rfa} and Forbes and Zhitnitsky \cite{Forbes:2008uf}, these A$\bar{\text{Q}}$Ns carry a (anti)baryon charge |B| $\approx 10^3 \div 10^{33}$. These values of the baryon number are typically constrained by the expected dark matter (DM) flux, and thus, the A$\bar{\text{Q}}$N flux.

The A$\bar{\text{Q}}$Ns made of antimatter are capable of releasing a significant amount of energy when they enter Earth’s atmosphere and annihilation processes start to occur between antimatter hidden in the form of A$\bar{\text{Q}}$Ns and the atmospheric or other materials such as rocks. In accord with the equation for the energy loss derived by De Rujula and Glashow \cite{DeRujula:1984axn}, and with the explicit form for the case of A$\bar{\text{Q}}$Ns:

\begin{equation}
    -\frac{dE}{dx} = \begin{cases}
\sigma_{\text{AQN}}\rho(x) v_{\text{AQN}}^2(L) & \text{if } v_{\text{AQN}}(L) \geq \sqrt{\frac{\varepsilon}{\rho}} \\
\varepsilon \sigma_{\text{AQN}} & \text{if } v_{\text{AQN}}(L) < \sqrt{\frac{\varepsilon}{\rho}}
\end{cases}
\end{equation}
where $\sigma_{\mathrm{AQN}}$ is the cross-section associated with the area of the nugget, and $\rho(x)$ is the density of the medium. 
For a well-defined distance $L$, 
\begin{equation}
    v_{\text{AQN}}(L) = v_{\text{AQN}}(0) \exp \left( -\frac{\sigma_{\text{AQN}}}{M_{\text{AQN}}} \int_0^L \rho \, dx \right).
\end{equation}

The equation of the energy loss breaks down at low velocity when the force generated by the particle becomes equal or lower than the force by which the material is confined. For velocities below $\sqrt{\varepsilon/\rho}$ the energy loss decreases with a constant rate and is brought to zero.
The variation of the velocity of nugget with distance can be considered as $v_{AQN}(L)$, where for the cross section can be considered geometric formula $\sigma_{AQN}=\pi R^2_{AQN}$ and $R_{AQN}=10^{-7}\left( \frac{B}{10^{24}} \right)^{1/3}$ [m].

A$\Bar{\mathrm{Q}}$Ns are highly dense objects with a radius on the order of µm. Thus, when such an object travels through matter the temperature should locally increase. In the lithosphere, the integrity of rocks persists up to the energy density $\varepsilon=6.2102$ eV/cm$^3$. Because most of the major rocks and minerals have very similar densities, around 2.6 to 3.0 g/cm$^3$, for velocities of nuggets below the value of 188 m/s, the energy loss is independent of their motion. At rest, nuggets will accumulate in the Earth’s crust and annihilate.

In the atmosphere, the parameters of interest vary strongly, depending on a multitude of factors. The atmosphere composition can be approximated considering only molecular nitrogen which represents nearly 78\% of the atmosphere concentration and oxygen 21\%. Molecular nitrogen has a triple bond between the two atoms, one sigma bond, and two pi bonds. This bond is very strong and requires 941 kJ/mol of energy to break and is only 495 kJ/mol for molecular oxygen. Considering an average density for the atmosphere around 0.657 kg/m$^3$ (as a mean value between sea level and and altitude of $2 \times 10^4$m), the energy loss is proportional with $v^2$ up to approximately 6.4 m/s. The average variation of air pressure, temperature and density with altitude (the standard atmosphere and the mean values of the parameters) are available in \cite{engineeringtoolbox}.

Initially, the A$\Bar{\mathrm{Q}}$N is neutral electric. The A$\Bar{\mathrm{Q}}$Ns made of antimatter are capable to release a significant amount of energy when they enter in the atmosphere and the Earth’s crust. In accord with \cite{Zhitnitsky_2021}, the binding energy of positrons is $E_{\mathrm{bound}} \sim$ keV. A large number of weakly bound positrons get excited and can leave the system.  If an electric field is produced in the vicinity of A$\Bar{\mathrm{Q}}$N through polarization effects, then positrons can be accelerated up to 10 MeV. As a result of these processes, A$\Bar{\mathrm{Q}}$N acquires a negative electric charge and supplementary will ionize the medium as a projectile with a mass $M \approx m_p B$.

In the work cited above, Ariel Zhitnitski invokes arguments related to the emission by A$\Bar{\mathrm{Q}}$N of axions and their expected characteristics.  The total energy of an A$\Bar{\mathrm{Q}}$N finds its equilibrium minimum when the axion domain wall contributes about 1/3 of its total mass and does not emit axions. When the annihilation processes start, the equilibrium state breaks because the mass of the A$\Bar{\mathrm{Q}}$N decreases, thus reducing its size. The surrounding domain wall starts to oscillate, generating excitation modes and emitting axions with a typical velocity of $v_{\text{axion}} \approx 0.6c$.
Unfortunately, the details of the annihilation process of baryons or nuclei with antiquark matter in the colour superconducting phase (2CS or CLF state) are not known. As a simple approximation, in order to estimate the particles generated after the annihilation process and their energy distributions, we supposed that the probability for the interaction of the antiquark core with nucleons is similar to the antiproton cross-section on nuclear matter. The mechanism in which antiprotons annihilate in interaction with nuclei was explained by Egidy \cite{VonEgidy:1987mz}, Richard \cite{Richard:2019dic} or Amsler and Myhrer \cite{amsler1991low}. In a standard scenario a primary annihilation produces mesons, and some of them penetrate the nucleus, giving rise to a variety of other phenomena: pion production, nucleon emission, internal excitation, etc. Complete and detailed experimental data exist in the classical paper of Chamberlain et al. \cite{PhysRev.113.1615}. 

The main experimental characteristics of the particles in the final states are: i) The number of pions (as average at rest and in flight): $ \langle N_{\pi} \rangle= 5.36 \pm 0.3$, with average total energy (at rest and in flight): $\langle E_{\pi} \rangle= 350 \pm 18 \mathrm{MeV}/ \pi$; ii)  Out of these pions, 1.3 and 1.9 interact with the nucleus at rest and in flight respectively, giving rise to nuclear excitation and nucleon emission; iii) 0.4 of interacting pions are inelastically scattered and the effect is a degradation of the primary pion energy to $\langle E_{\pi} \rangle= 339 \pm 18 \mathrm{MeV}$; iv)  An average number of $1.6 \pm 0.1$ of the pions produced in the annihilation interacts with the nucleus in which the annihilation occurs, with the effect of nuclear excitation and nucleon emission. The average number of protons emitted per annihilation is $\langle N_B \rangle= 4.1 \pm 0.3$ and the corresponding total average energy release in protons and neutrons is $\langle \Sigma E_B \rangle= 490 \pm 40 \mathrm{MeV}$. 

In the laboratory frame, from $\pi_0 \to \gamma + \gamma$ the energies of photons are in the energy range $0 \le E_{\gamma} \le E_{\pi^0}$ with a flat distribution. From the decay of charged pions, $\pi \to \mu + \nu$, the energy of muon is $0.58 E_{\pi} \le E_{\mu} \le E_{\pi}$. Since the annihilation processes occur within nuclei, nuclear effects may alter these results. However, the similarity between values at rest and in flight suggests minimal differences. 

If the annihilation takes place in the nucleus, in accord with the description of Plendl and co-workers \cite{HSPlendl_1993}, the principal steps that are assumed to take place just before, during and after the intranuclear cascade (INC) caused by annihilation are the following: a) capture of antimatter structure into high-n atomic orbit;  b) cascade to low-n atomic orbit as intermediate steps from annihilation; c) Mesons: $K$, $\eta$, $\rho$, $\omega$, $\pi$, between 2 up to 8, with energies: 20 to 600 MeV/$\pi$ are produced; d) Direct emission of kaons, pions from annihilation; e)  Other successive processes: multi-pion induced INC of $\pi N$, $\pi NN$, $NN$, energetic $n$, $p$ from INC coalescence and emission of $d$, $t$, $^3$He, $^4$He or from pre-equilibrium, multifragmentation, and others take place. The time scale for all these processes is between 0 s (direct annihilation) up to $10^{-18} \div 10^{-17}$ s. 

Processes initiated in A$\Bar{\mathrm{Q}}$N present critical differences compared with nucleon – antinucleon annihilation. Depending on the parameters, the inside of the A$\Bar{\mathrm{Q}}$N will exist in a color superconducting phase (CS or CFL) with a gap parameter $\Delta \approx 100$ MeV, while the outside will consist of a free Fermi gas of positrons to maintain neutrality. In the CFL phase, colored quark matter and gluons at high densities can be in excited states. They are strongly interacting quasi-particles and the energy will ultimately be transferred to lower energy degrees of freedom which are the Nambu-Goldstone (NG) bosons. The NG bosons are collective excitations of a diquark condensate. More details from the theoretical point of view and concrete equations for masses calculations and the dispersion relations are given in the papers of Alford and collaborators \cite{Alford_2008} and Lawson and Zhitnisky \cite{PhysRevD.95.063521}. For these states, peculiarities include an inverted mass spectrum, with kaons being lighter than pions.

In this article we are interested in determining the energy deposited by the A$\Bar{\mathrm{Q}}$N following its annihilation in rocks in order to find a direct detection method through absorbed dose measurements. Zhitnitsky \cite{Zhitnitsky_2021} and Budker et al. \cite{Budker_2022} estimated the number of frontal collisions of the environmental molecules with an A$\Bar{\mathrm{Q}}$N per unit time, the energy rate deposited by A$\Bar{\mathrm{Q}}$Ns as a result of annihilation processes and the energy transfer per unit distance in the material of the environment.
Unfortunately, in the absence of relevant experimental results, the energy transfer following the annihilation process from the nuclear state of colored superconductivity to that of normal nuclear matter, along with the kinematic properties of the resulting particles are not fully understood.

In order to estimate the annihilation probability for the incident protons on the A$\Bar{\mathrm{Q}}$N core, it is assumed that the typical cross section is of the order of antiproton cross section on the nuclear matter and a value of $\sigma \approx$ 0.4 b was considered \cite{Flambaum_2021,PhysRevD.105.123011,PhysRevD.106.023006}.  This annihilation process is considered near the surface of the A$\Bar{\mathrm{Q}}$N and different estimations of the energies of different particles produced after annihilations are given. In a recent paper, the same authors \cite{flambaum2024manifestation} introduced a series of constraints in the case of annihilations, considering the absorption of energy after these processes through strong and electromagnetic interactions that prevent weak disintegrations with neutrino emission. The authors estimated that a suppression factor $<3.3 \times 10^{-4}$ must be considered for the emitted mesons.
The numerical results that we will present below will use these assumptions.

\section{Basic concepts for detection of the A$\Bar{\mathrm{Q}}$Ns}

Because the fluxes of A$\Bar{\mathrm{Q}}$Ns are unknown and not ambiguously observed until now, the conservative assumption is that the interaction rates, if they exist, are small. Therefore, massive detectors sensitive to these particles and long exposure times are required, or it is necessary to use nonstandard techniques for detection. Large classes of geological minerals present sensitivity to energies deposed in different interaction processes with projectile particles. An additional requirement is that the signals produced in the form of latent signals should be stable for as long as possible, depending on the ambient conditions. For different materials, due to the link of the luminescence process with the deposited energy, this is an excitation process used for dosimetric purposes. The predicted lifetimes \cite{POLYMERIS2006207}, for storage at 15 $^{\circ}$C is of the order of $10^8$ yr, and decrease exponentially with increasing temperature.

Fluorite (CaF$_2$) is commercially known as fluorspar. The deposits of fluorite occur in a variety of geologic environments throughout the globe - see for example Figure 1 from reference \cite{magotra2017new}. Silicate minerals are rock-forming minerals made up of silicate groups. These minerals represent up to approximately 90 percent of Earth’s crust. For the present study especially SiO$_2$ is important. The U.S. Mineral Resources Data System (MRDS) \cite{usgs_mrds} offers comprehensive maps and precise location data of silicon dioxide (SiO$_2$) and calcium fluoride (CaF$_2$) deposits.


Sedimentary quartz (SiO$_2$), natural calcium fluoride (CaF$_2$) and feldspars (x[AlSi$_3$O$_8$] with x= K, Na, Ca) present luminescence properties and are large distributed in the Earth. Minerals, exposed to different types of radiation since the time of their formation, have accumulated over time doses of radiation in the form of latent signals, which can be extracted by appropriate physical methods. In the case of luminescence signals, the "reading" of this information can be extracted through the thermoluminescence (TL) or Optically Stimulated Luminescence (OSL) techniques. 
Conventionally, TL measurements are made by recording light emission during heating sample. Much more information may be gained by monitoring the details of the emission spectrum during thermoluminescence. TL spectra of minerals exhibit changes as a result of crystal purity, radiation dose, dose rate and thermal history. \cite{rendell1993thermoluminescence}.

If the dose accumulation is proportional to the deposited energy and there are no fading effects, then the read dose will be proportional to the total energy. In the case of rare events, in principle, the total integrated dose is the sum of contributions due to the cosmic background or from other sources and singular contributions, with random spatial locations associated with axion antiquark nuggets type events.
Different components relevant for the calculation of dose rate for luminescence include: cosmic rays from space (protons, heavy ions, electrons, muons, photons, etc.), which decrease with depth; external radiation flux from neighboring grains, such as $^{40}$K, $^{232}$Th, and $^{238}$U chains (including $\alpha$, $\beta$, and $\gamma$ radiation); and internal radiation, potentially from $^{40}$K and other sources.

This background, as paleodose $D_b$ [Gy], is calculated as the integrated luminescence [yr] multiplied by the dose rate [Gy/yr]. If a singular annihilation event of an A$\Bar{\mathrm{Q}}$N occurs in a region with mineral deposits of interest (such as feldspars, quartz, calcite, etc.) at a certain depth, an overdose (D$_{\text{AQN}}$) with a certain spatial distribution and approximately spherical symmetry will be generated, so that:
\begin{equation}
D_{AQN}=D- D_b.    
\end{equation}

\section{Numerical results and predicted doses produced by the annihilation of A$\Bar{\mathrm{Q}}$Ns}

\subsection{Energy loss of the A$\Bar{\mathrm{Q}}$N in air and rocks}

The energy produced by annihilation inside the A$\Bar{\mathrm{Q}}$N when it travels the atmosphere or underground per unit length can be expressed as follows \cite{Budker_2022}:




\begin{equation}
    -\frac{dE}{dx} = \begin{cases}
 \approx 10^4 \cdot \kappa \left( \frac{B}{10^{25}} \right)^{2/3} \left( \frac{n_{\mathrm{air}}}{10^{21} \mathrm{cm}^{-3}} \right) \frac{\mathrm{J}}{{\mathrm{m}}} \text{ for atmosphere},\\
 \approx 10^7 \cdot \kappa \left( \frac{B}{10^{25}} \right)^{2/3} \left( \frac{n_{\mathrm{rock}}}{10^{24} \mathrm{cm}^{-3}} \right) \frac{\mathrm{J}}{{\mathrm{m}}} \text{ for rocks},
\end{cases}
\end{equation}
where $n_{\mathrm{air}}$ is the total number of nucleons in atoms such that $\rho_{\mathrm{air}} = n_{\mathrm{air}} m_p$, thus $n_{\mathrm{air}} = 7.7 \times 10^{21} \mathrm{cm}^{-3}$ and $n_{\mathrm{rock}}$ is the total number of nucleons in atoms such that $\rho_{\mathrm{rock}} = n_{\mathrm{rock}} m_p$ , thus $n_{\mathrm{rock}} = 1.8 \times 10^{24} \mathrm{cm}^{-3}$. $\kappa$ represents a parameter that acknowledges the phenomenon where not all matter striking the nugget will undergo annihilation, and not all of the energy released by an annihilation event will be fully thermalized within the nuggets. For instance, a portion of the energy will be emitted in the form of axions and neutrinos.

Considering $\kappa \approx 1$, the energy released in the atmosphere is $\sim 7.7$ kJ/m and the energy released in rocks is $\sim 18 $ MJ/m. Increasing the density of the surrounding material significantly amplifies the released annihilation energy. In the underground case, if one removes the low-energy positrons from the electrosphere, an additional suppression factor $\xi \sim 10^{-2}$ and even much smaller has to be introduced in the estimation of the energy loss via annihilation processes.

The internal temperature of the A$\Bar{\mathrm{Q}}$N propagating in the Earth’s atmosphere can be expressed as \cite{zhitnitsky2024mysterious}:
\begin{equation}
    T_{\mathrm{atm}} \approx 40 \mathrm{keV} \cdot \left( \frac{n_{\mathrm{atm}}}{10^{21} \mathrm{cm}^{-3}} \right) ^{4/17} \kappa^{4/17},
\end{equation}
and the electric charge of the A$\Bar{\mathrm{Q}}$N can be written as a function of the temperature $T$ as \cite{Zhitnitsky_2021}:
\begin{equation}
    Q \approx \frac{4 \pi R^2}{\sqrt{2 \pi \alpha}} (m_e T) \frac{T}{m_e}^{1/4}
\end{equation}

In case of solids the temperature is $T \approx (100-200)$ keV, thus the electric charge of the A$\Bar{\mathrm{Q}}$N becomes $\approx (7.8 \times 10^8 \div 1.86 \times 10^9) |\mathrm{e}|$, thus the energy loss via ionization is $\approx 10^{16}$ GeV/cm. The collision of A$\Bar{\mathrm{Q}}$Ns with dense media such as rocks should release a huge amount of energy, on the order of hundreds of MJ/m, equivalent to several kg of TNT in accord with Table 1 from Ref. \cite{flambaum2024manifestation}.


The rate at which antimatter quark nuggets (A$\Bar{\mathrm{Q}}$Ns) hit Earth's surface is very low, and can be calculated as \cite{Liang:2019lya}: 

\begin{equation}
\varnothing_{\text{AQN}}^{\text{Earth}} = \frac{\langle \dot{N} \rangle}{4\pi R_{\text{Earth}}^2} = 0.4 \times \frac{10^{24}}{\langle B \rangle} \frac{\rho_{\text{DM}}}{0.3 \, \text{GeV/cm}^3} \frac{\langle v_{\text{AQN}} \rangle}{220 \, \text{km/s}} \left[ \frac{1}{\text{km}^2 \, \text{yr}} \right]
\end{equation}

This indicates that lighter A$\Bar{\mathrm{Q}}$Ns tend to have a higher overall flux. Given that the average speed of local dark matter is \(\langle v_{\text{AQN}} \rangle\) and the density \(\rho_{\text{DM}}\) is approximately 0.3-0.4 GeV/cm\(^3\) \cite{deSalas:2019pee}, the estimated flux of AQNs hitting Earth (\(\varnothing_{\text{AQN}}^{\text{Earth}}\)) is:

\begin{equation}
\varnothing_{\text{AQN}}^{\text{Earth}} =
\begin{cases}
2 \times 10^{12} \text{ AQNs/year,} & \text{for } \langle B \rangle = 10^{20}, \\
2 \times 10^{7} \text{ ~~AQNs/year,} & \text{for } \langle B \rangle = 10^{25}.
\end{cases}
\end{equation}


To simulate the expected dose, we have used the FLUKA 4-3.4 code \cite{ahdida2022new,battistoni2015overview} alongside the Flair graphical interface \cite{vlachoudis2009flair}. FLUKA code can simulate nuclear reactions, produce secondary particles and model advanced geometries in order to ensure accurate and reliable assessment of the deposited doses in different types of environments. The transport of charged particles is carried out using the Multiple Coulomb scattering algorithm, with an optional single scattering method also available. The electromagnetic physics models in FLUKA describe continuous energy losses of charged particles, energy loss straggling, delta-ray production, and multiple Coulomb scattering. 

\begin{figure}[h!]
  \centering
  \begin{subfigure}[b]{0.45\textwidth}
    \centering
    \includegraphics[width=\linewidth]{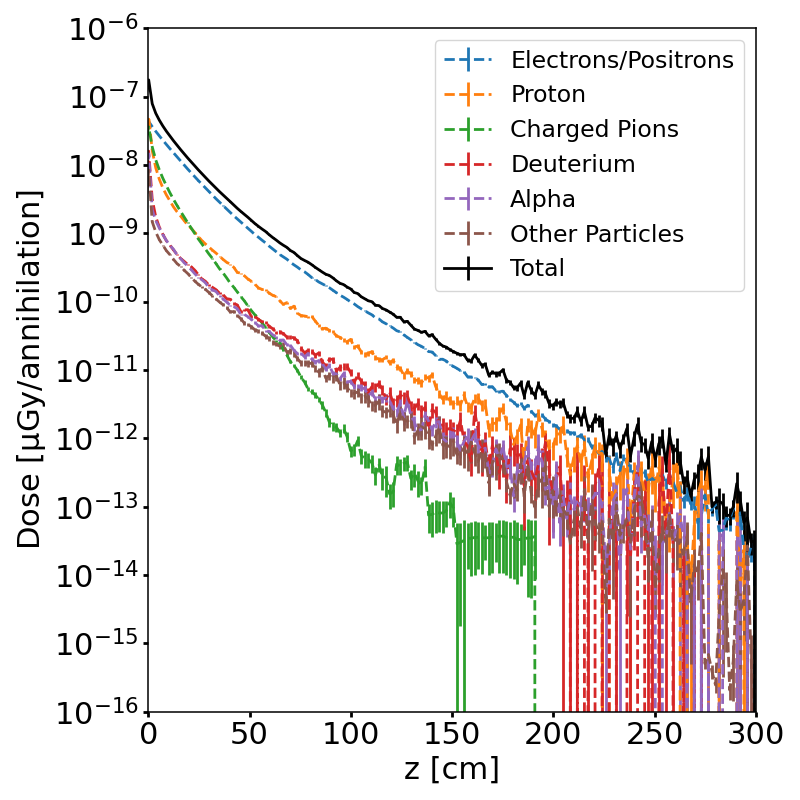}
    \label{CaF2_dose}
  \end{subfigure}
  \hfill
  \begin{subfigure}[b]{0.45\textwidth}
    \centering
    \includegraphics[width=\linewidth]{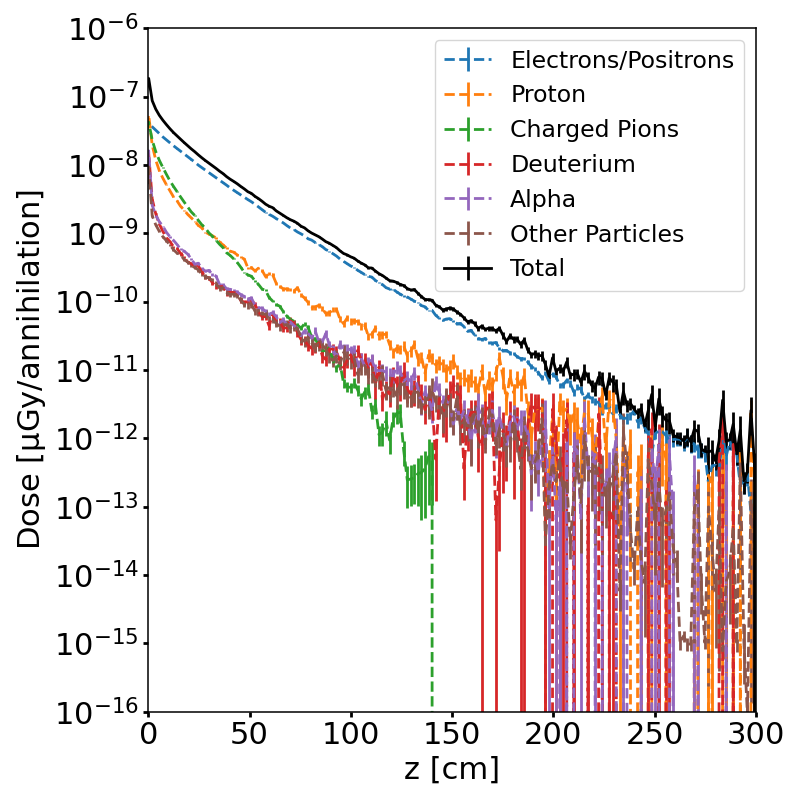}
    \label{SiO2_dose}
  \end{subfigure}
  \caption{The dose deposited by \(e^+/e^-\), \(p\), \(\pi^+/\pi^-\), \(d\), \(\alpha\) in one annihilation process in CaF$_2$ (left) and SiO$_2$ (right) as a function of the distance from the annihilation point. The difference between the total deposited dose and the contribution of these particles is labeled as "Other particles", i.e., \(n\), \(\mu\), \(\tau\), \(K\), etc. The suppression of the emission of some particles, particularly pions, was not taken into account.}
  \label{CaF2_SiO2_dose}
\end{figure}

\begin{figure}[h!]
  \centering
  \includegraphics[width=.7\linewidth]{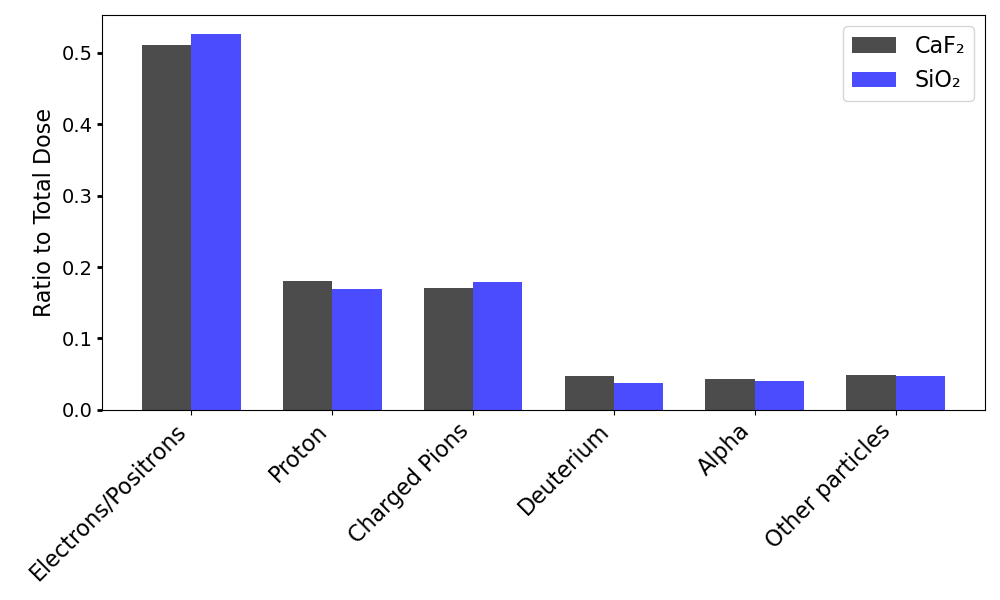} 
  \caption{Fraction of dose generated by each type of particle in CaF$_2$ and SiO$_2$ for each type of particle produced in the antiproton annihilation event. The main contributions are from electrons and positrons, along with protons and charged pions. The suppression of the emission of some particles, particularly pions, was not taken into account.} 
  \label{CaF2_SiO2_dose-ratio} 
\end{figure}

The spatial distribution of particles produced in $\bar{\mathrm{p}}$ annihilation in CaF$_2$ and SiO$_2$ is spherically symmetric. 
Figure \ref{CaF2_SiO2_dose} represents the dose deposited by $e^+/e^-, p, \pi^+/\pi^-, d, \alpha$ in one annihilation process in CaF$_2$ and SiO$_2$ respectively with respect to the distance from the annihilation point. The difference between the total deposited dose and the contribution of these particles is labeled as "Other particles", i.e. $n, \mu, t, K$ etc. The ratio between the deposited dose by each type of particle and the total dose is represented in Figure \ref{CaF2_SiO2_dose-ratio}. It can be seen that the main contribution is from electrons and positrons, along with protons and charged pions.

\begin{figure}[h!]
  \centering
  \includegraphics[width=.65\linewidth]{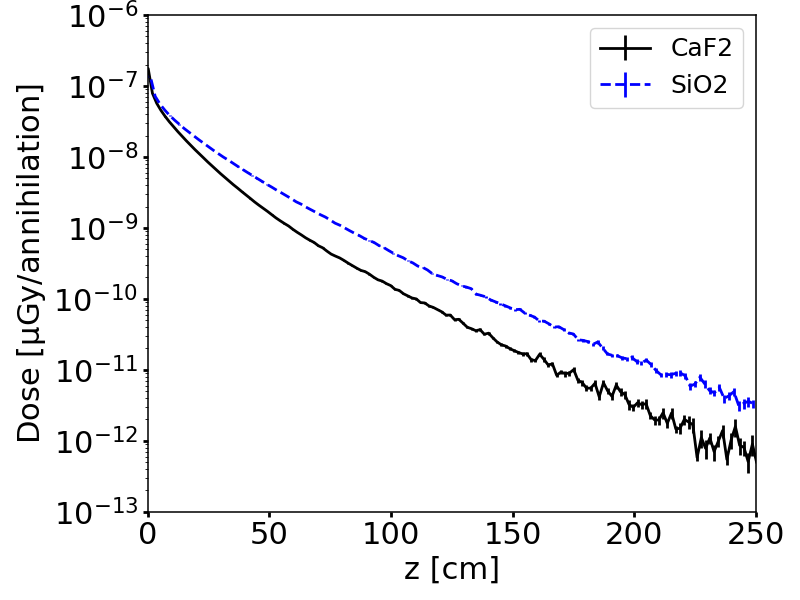} 
  \caption{A comparison between the total dose generated by a single antiproton annihilation in CaF$_2$ and SiO$_2$, as a function of the z-axis. In order to obtain the total deposited dose, these values should be multiplied by the number of annihilation processes.} 
  \label{CaF2_SiO2_total-dose} 
\end{figure}

A comparison between the total doses generated by particles from a single annihilation process in each medium, as a function of the z-axis, is shown in Figure \ref{CaF2_SiO2_total-dose}.

In order to get the total deposited dose, these values should be multiplied by the number of annihilation processes. Luminescence dosimetry has a high level of sensitivity, with remarkably low detection limits for radiation doses. The smallest dose recorded using a natural dosimeter with TL technology is 1 \textmu Gy, achieved with calcium fluoride (CaF2:N). In contrast, the Lowest Detectable Dose Limit (LDDL) for sedimentary quartz dosimeters is 1 mGy, although in some cases, this limit can be reduced by one order of magnitude \cite{mckeever1995thermoluminescence,aitken1998introduction,POLYMERIS2006207}. 

Analysing the energies deposited in the two minerals by all types particles following the annihilation, we reached the following conclusions: a) Over 50\% of the deposited energy is due to electrons/positrons, over 20\% is due to pions, and the remainder is due to other particles. b) Working under conservative assumptions, and including the suppression of the emission of some particles, particularly pions, the doses contributed by electrons/positrons remain unaffected. c) If we consider only the doses due to electrons and positrons, then, at a distance of 1 meter from the place of annihilation, the value obtained will exceed the Lowest Detectable Dose Limit in CaF$_2$ if the number of annihilated protons is greater than $10^{10}$ and in SiO$_2$ if more than $10^{12}$ annihilations occur.

\subsection{Numerical values for the contribution of the radioactive background to dose}

The cosmic rays interact with the nuclei of atmospheric constituents, producing a cascade of interactions and secondary reaction products that contribute to cosmic ray exposure that decreases in intensity with depth in the atmosphere, from the stratosphere to the ground level. 
The primary types of radiation that originate in outer space and impinge on the top of the earth’s atmosphere consist of 87\% protons, 11\% $\alpha$ particles, about 1\% nuclei of atomic number between 4 (beryllium) and 26 (iron), and about 1\% electrons of very high energy. They originate outside the solar system, and only a small fraction is normally of solar origin; however, the solar component becomes significant during solar flares, which follows an 11-year cycle. The interactions of the primary particles with atmospheric nuclei produce electrons, $\gamma$ rays, neutrons and mesons. At the ground level, the dominant component of the cosmic ray field consists of neutrinos and muons with energy mainly between 1 and 20 GeV \cite{cinelli2017european}.

A world-wide contour map of the cosmic-ray dose rates at the ground level is available at \cite{expacs_dosemap}, for an interval of dose rate between 20 to 600 nSv/h (or equivalent 0.18 to 5.26 mSv/yr). The values are  calculated in the frame of the model PARMA coupled with the global relief data ETOPO2v2 which is provided by National Geophysical Data Center.

In the absence of direct information related to the dose rates produced by the abundances of uranium, thorium, potassium and their descendants in minerals and rocks, we will investigate the abundances of these elements in the different classes of materials, predominantly those with low contents of radioactive elements. Most of the information presented here are from the Ref. \cite{rocks_activity}. For uranium, igneous rocks as basalts and other mafic rocks or ultramafic are characterized by abundances between 0.001–1 ppm, intermediate rocks 1–6 ppm, Granites and rhyolites 2–50 ppm or syenites and phonolites 0.1–26 ppm. Between sedimentary rocks shales, clays, mudrocks are 1–5 ppm, sandstones 0.5–4 ppm, limestones, dolomites <0.1–9 ppm. The metamorphic rocks are characterized by abundances <1–5 ppm. The thorium is a relatively rare element. For basalts and other mafic rocks, ultramafic rocks and limestones, dolomites the abundances are in the range <0.05–3 - 4 ppm, the lower abundances. 

Potassium abundance in igneous rocks are the lower values in ultramafic rocks (<10 ppm–1\%) and basalts and other mafic rocks (1–2)\%. In sedimentary rocks in pure gypsum potassium not exist, and rocks of interest are shales, clays, mudrocks, limestones and dolomites with abundances <0.01–7.1\%.
In order to look for singular events of interaction of A$\Bar{\mathrm{Q}}$N with these rocks, first of all the complete or partial annihilation of the A$\Bar{\mathrm{Q}}$N core, the rocks/minerals in which these processes must be searched for, must be characterized by the lowest abundance of uranium, thorium and potassium, the doses accumulated from cosmic radiation should be as low as possible, and the temperature of the environment constant and as low as possible throughout the year.


\section{Summary and conclusions}

The paper investigates the possibility that A$\bar{\mathrm{Q}}$Ns, potential exotic particles associated with dark matter, could be detected using natural mineral deposits (CaF$_2$ and SiO$_2$) by exploiting TL/OSL techniques. The proposed approach involves identifying regions with excess dose in these materials. Because A$\Bar{\mathrm{Q}}$Ns are currently hypothetical particles, a series of simplifying hypotheses have been considered for their interactions with matter. If this region is characterized by spherical symmetry and has a spatial extension predicted by calculations, it is plausible to attribute these signals to A$\bar{\mathrm{Q}}$N annihilations. Given a radioactive background kept as low as possible, and a minimum detectable dose limit for TL/OSL in both materials in accordance with values reported in the literature, A$\bar{\mathrm{Q}}$Ns can be detected if at least $10^{10}$ protons annihilate in CaF$_2$ and $10^{12}$ in SiO$_2$ respectively.

\section*{Acknowledgments}
We express our deep gratitude to Professor Konstantin Zioutas for inviting and encouraging us to publish this article, as well as for the relevant scientific discussions related to the components of dark matter and its interactions in the universe. His comments and new ideas were extremely stimulating for our work. We also express our gratitude to Professor Ariel Zhitnitsky for his comments on certain aspects of the mechanisms of the interactions of nuggets with matter. This work was performed with the financial support of the Romanian Program PNCDI III, Programme 5, Module 5.2 CERN-RO, under contract no. 04/2022.

\printbibliography

@article{Witten:1984rs,
    author = "Witten, Edward",
    title = "{Cosmic Separation of Phases}",
    reportNumber = "PRINT-84-0400 (IAS,PRINCETON)",
    doi = "10.1103/PhysRevD.30.272",
    journal = "Phys. Rev. D",
    volume = "30",
    pages = "272--285",
    year = "1984"
}

@article{Zhitnitsky:2002qa,
    author = "Zhitnitsky, Ariel R.",
    title = "{'Nonbaryonic' dark matter as baryonic color superconductor}",
    eprint = "hep-ph/0202161",
    archivePrefix = "arXiv",
    doi = "10.1088/1475-7516/2003/10/010",
    journal = "JCAP",
    volume = "10",
    pages = "010",
    year = "2003"
}

@article{DeRujula:1984axn,
    author = "De Rujula, A. and Glashow, S. L.",
    title = "{Nuclearites: A Novel Form of Cosmic Radiation}",
    reportNumber = "HUTP-84-A057",
    doi = "10.1038/312734a0",
    journal = "Nature",
    volume = "312",
    pages = "734--737",
    year = "1984"
}

@article{SinghSidhu:2020cxw,
    author = "Singh Sidhu, Jagjit and Scherrer, Robert J. and Starkman, Glenn",
    title = "{Antimatter as macroscopic dark matter}",
    eprint = "2006.01200",
    archivePrefix = "arXiv",
    primaryClass = "astro-ph.CO",
    doi = "10.1016/j.physletb.2020.135574",
    journal = "Phys. Lett. B",
    volume = "807",
    pages = "135574",
    year = "2020"
}

@article{BAUM2023101245,
title = {Mineral detection of neutrinos and dark matter. A whitepaper},
journal = {Physics of the Dark Universe},
volume = {41},
pages = {101245},
year = {2023},
issn = {2212-6864},
doi = {https://doi.org/10.1016/j.dark.2023.101245},
url = {https://www.sciencedirect.com/science/article/pii/S2212686423000791},
author = {Sebastian Baum and Patrick Stengel and Natsue Abe and Javier F. Acevedo and Gabriela R. Araujo and Yoshihiro Asahara and Frank Avignone and Levente Balogh and Laura Baudis and Yilda Boukhtouchen and Joseph Bramante and Pieter Alexander Breur and Lorenzo Caccianiga and Francesco Capozzi and Juan I. Collar and Reza Ebadi and Thomas Edwards and Klaus Eitel and Alexey Elykov and Rodney C. Ewing and Katherine Freese and Audrey Fung and Claudio Galelli and Ulrich A. Glasmacher and Arianna Gleason and Noriko Hasebe and Shigenobu Hirose and Shunsaku Horiuchi and Yasushi Hoshino and Patrick Huber and Yuki Ido and Yohei Igami and Norito Ishikawa and Yoshitaka Itow and Takashi Kamiyama and Takenori Kato and Bradley J. Kavanagh and Yoji Kawamura and Shingo Kazama and Christopher J. Kenney and Ben Kilminster and Yui Kouketsu and Yukiko Kozaka and Noah A. Kurinsky and Matthew Leybourne and Thalles Lucas and William F. McDonough and Mason C. Marshall and Jose Maria Mateos and Anubhav Mathur and Katsuyoshi Michibayashi and Sharlotte Mkhonto and Kohta Murase and Tatsuhiro Naka and Kenji Oguni and Surjeet Rajendran and Hitoshi Sakane and Paola Sala and Kate Scholberg and Ingrida Semenec and Takuya Shiraishi and Joshua Spitz and Kai Sun and Katsuhiko Suzuki and Erwin H. Tanin and Aaron Vincent and Nikita Vladimirov and Ronald L. Walsworth and Hiroko Watanabe},

}

@article{POLYMERIS2006207,
title = {Minerals as Time-Integrating Luminescence Detectors for setting bounds on dark matter particle characteristics},
journal = {Nuclear Instruments and Methods in Physics Research Section A: Accelerators, Spectrometers, Detectors and Associated Equipment},
volume = {562},
number = {1},
pages = {207-213},
year = {2006},
issn = {0168-9002},
doi = {https://doi.org/10.1016/j.nima.2006.01.126},
url = {https://www.sciencedirect.com/science/article/pii/S0168900206002282},
author = {G.S. Polymeris and G. Kitis and A.K. Liolios and N.C. Tsirliganis and K. Zioutas},
}

@inproceedings{Lawson:2013bya,
    author = "Lawson, Kyle and Zhitnitsky, Ariel R.",
    title = "{Quark (Anti) Nugget Dark Matter}",
    booktitle = "{Snowmass 2013}: {Snowmass on the Mississippi}",
    eprint = "1305.6318",
    archivePrefix = "arXiv",
    primaryClass = "astro-ph.CO",
    month = "5",
    year = "2013"
}

@article{Gorham:2015rfa,
    author = "Gorham, P. W. and Rotter, B. J.",
    title = "{Stringent neutrino flux constraints on antiquark nugget dark matter}",
    eprint = "1507.03545",
    archivePrefix = "arXiv",
    primaryClass = "astro-ph.CO",
    doi = "10.1103/PhysRevD.95.103002",
    journal = "Phys. Rev. D",
    volume = "95",
    number = "10",
    pages = "103002",
    year = "2017"
}

@misc{engineeringtoolbox,
  title = {Standard Atmosphere},
  author = {Engineering Toolbox},
  year = {2024},
  url = {https://www.engineeringtoolbox.com/standard-atmosphere-d_604.html}
}

@article{Zhitnitsky_2021,
   title={Axion quark nuggets. Dark matter and matter–antimatter asymmetry: Theory, observations and future experiments},
   volume={36},
   ISSN={1793-6632},
   url={http://dx.doi.org/10.1142/S0217732321300172},
   DOI={10.1142/s0217732321300172},
   number={18},
   journal={Modern Physics Letters A},
   publisher={World Scientific Pub Co Pte Lt},
   author={Zhitnitsky, Ariel},
   year={2021},
   month=may, pages={2130017} }

@article{PhysRev.113.1615,
  title = {Antiproton-Nucleon Annihilation Process. II},
  author = {Chamberlain, Owen and Goldhaber, Gerson and Jauneau, Louis and Kalogeropoulos, Theodore and Segr\`e, Emilio and Silberberg, Rein},
  journal = {Phys. Rev.},
  volume = {113},
  issue = {6},
  pages = {1615--1634},
  numpages = {0},
  year = {1959},
  month = {Mar},
  publisher = {American Physical Society},
  doi = {10.1103/PhysRev.113.1615},
  url = {https://link.aps.org/doi/10.1103/PhysRev.113.1615}
}

@article{VonEgidy:1987mz,
    author = "Von Egidy, T.",
    title = "{Interaction and Annihilation of Anti-protons and Nuclei}",
    doi = "10.1038/328773a0",
    journal = "Nature",
    volume = "328",
    pages = "773--778",
    year = "1987"
}

@article{Richard:2019dic,
    author = "Richard, Jean-Marc",
    title = "{Antiproton physics}",
    eprint = "1912.07385",
    archivePrefix = "arXiv",
    primaryClass = "nucl-th",
    doi = "10.3389/fphy.2020.00006",
    journal = "Front. in Phys.",
    volume = "8",
    pages = "6",
    year = "2020"
}

@article{rendell1993thermoluminescence,
  title={Thermoluminescence spectra of minerals},
  author={Rendell, HM and Khanlary, M-R and Townsend, PD and Calder{\'o}n, T and Luff, BJ},
  journal={Mineralogical Magazine},
  volume={57},
  number={387},
  pages={217--222},
  year={1993},
  publisher={Cambridge University Press}
}

@article{cinelli2017european,
  title={European annual cosmic-ray dose: estimation of population exposure},
  author={Cinelli, Giorgia and Gruber, Valeria and De Felice, Luca and Bossew, Peter and Hernandez-Ceballos, Miguel Angel and Tollefsen, Tore and Mundigl, Stefan and De Cort, Marc},
  journal={Journal of maps},
  volume={13},
  number={2},
  pages={812--821},
  year={2017},
  publisher={Taylor \& Francis}
}

@misc{rocks_activity,
   author = "MediaWiki",
   title = "OR/17/001 The distribution of natural radioactivity in rocks",
   year = "2019"
 }

@article{Budker_2022,
   title={Infrasonic, Acoustic and Seismic Waves Produced by the Axion Quark Nuggets},
   volume={14},
   ISSN={2073-8994},
   url={http://dx.doi.org/10.3390/sym14030459},
   DOI={10.3390/sym14030459},
   number={3},
   journal={Symmetry},
   publisher={MDPI AG},
   author={Budker, Dmitry and Flambaum, Victor V. and Zhitnitsky, Ariel},
   year={2022},
   month=feb, pages={459} }

@article{Liang:2019lya,
    author = "Liang, Xunyu and Mead, Alexander and Siddiqui, Md Shahriar Rahim and Van Waerbeke, Ludovic and Zhitnitsky, Ariel",
    title = "{Axion Quark Nugget Dark Matter: Time Modulations and Amplifications}",
    eprint = "1908.04675",
    archivePrefix = "arXiv",
    primaryClass = "astro-ph.CO",
    doi = "10.1103/PhysRevD.101.043512",
    journal = "Phys. Rev. D",
    volume = "101",
    number = "4",
    pages = "043512",
    year = "2020"
}

@article{deSalas:2019pee,
    author = "de Salas, P. F. and Malhan, K. and Freese, K. and Hattori, K. and Valluri, M.",
    title = "{On the estimation of the Local Dark Matter Density using the rotation curve of the Milky Way}",
    eprint = "1906.06133",
    archivePrefix = "arXiv",
    primaryClass = "astro-ph.GA",
    doi = "10.1088/1475-7516/2019/10/037",
    journal = "JCAP",
    volume = "10",
    pages = "037",
    year = "2019"
}

@article{ahdida2022new,
  title={New capabilities of the FLUKA multi-purpose code},
  author={Ahdida, C and Bozzato, D and Calzolari, D and Cerutti, F and Charitonidis, N and Cimmino, A and Coronetti, A and D’Alessandro, GL and Donadon Servelle, A and Esposito, LS and others},
  journal={Frontiers in Physics},
  volume={9},
  pages={788253},
  year={2022},
  publisher={Frontiers}
}

@article{battistoni2015overview,
  title={Overview of the FLUKA code},
  author={Battistoni, Giuseppe and Boehlen, Till and Cerutti, Francesco and Chin, Pik Wai and Esposito, Luigi Salvatore and Fass{\`o}, Alberto and Ferrari, Alfredo and Lechner, Anton and Empl, Anton and Mairani, Andrea and others},
  journal={Annals of Nuclear Energy},
  volume={82},
  pages={10--18},
  year={2015},
  publisher={Elsevier}
}

@inproceedings{vlachoudis2009flair,
  title={FLAIR: a powerful but user friendly graphical interface for FLUKA},
  author={Vlachoudis, Vasilis and others},
  booktitle={Proc. Int. Conf. on Mathematics, Computational Methods \& Reactor Physics (M\&C 2009), Saratoga Springs, New York},
  volume={176},
  year={2009}
}

@article{aitken1998introduction,
  title={An introduction to optical dating: Oxford University Press},
  author={Aitken, MJ},
  journal={New York},
  year={1998}
}

@article{mckeever1995thermoluminescence,
  title={Thermoluminescence dosimetry materials: properties and uses},
  author={McKeever, S WS and Moscovitch, Marko and Townsend, Peter David},
  year={1995}
}

@article{HSPlendl_1993,
doi = {10.1088/0031-8949/48/2/006},
url = {https://dx.doi.org/10.1088/0031-8949/48/2/006},
year = {1993},
month = {aug},
publisher = {},
volume = {48},
number = {2},
pages = {160},
author = {H S Plendl and  H Daniel and  T von Egidy and  T Haninger and  F S Hartmann and  P Hofmann and  Y S Kim and  H Machner and  G Riepe and  J Jastrzebski and  A Grabowska and  W Kurcewicz and  P Lubinski and  A Stolarz and  A S Botvina and  Ye S Golubeva and  A S Iljinov and  V G Nedorezov and  A S Sudov and  K Ziock},
title = {Antiproton-nucleus annihilation at rest},
journal = {Physica Scripta}
}

@misc{zhitnitsky2024mysterious,
      title={Mysterious anomalies in Earth's atmosphere and strongly interacting Dark Matter}, 
      author={Ariel Zhitnitsky},
      year={2024},
      eprint={2405.04635},
      archivePrefix={arXiv},
      primaryClass={hep-ph}
}

@article{Lazanu_2024,
doi = {10.1088/1475-7516/2024/05/014},
url = {https://dx.doi.org/10.1088/1475-7516/2024/05/014},
year = {2024},
month = {may},
publisher = {IOP Publishing},
volume = {2024},
number = {05},
pages = {014},
author = {I. Lazanu and M. Parvu},
title = {Exploring the detection of AQNs in large liquid detectors},
journal = {Journal of Cosmology and Astroparticle Physics}
}

@article{Alford_2008,
   title={Color superconductivity in dense quark matter},
   volume={80},
   ISSN={1539-0756},
   url={http://dx.doi.org/10.1103/RevModPhys.80.1455},
   DOI={10.1103/revmodphys.80.1455},
   number={4},
   journal={Reviews of Modern Physics},
   publisher={American Physical Society (APS)},
   author={Alford, Mark G. and Schmitt, Andreas and Rajagopal, Krishna and Schäfer, Thomas},
   year={2008},
   month=nov, pages={1455–1515} }

@article{PhysRevD.95.063521,
  title = {Solar neutrino spectrum of quark nugget dark matter},
  author = {Lawson, K. and Zhitnitsky, A. R.},
  journal = {Phys. Rev. D},
  volume = {95},
  issue = {6},
  pages = {063521},
  numpages = {11},
  year = {2017},
  month = {Mar},
  publisher = {American Physical Society},
  doi = {10.1103/PhysRevD.95.063521},
  url = {https://link.aps.org/doi/10.1103/PhysRevD.95.063521}
}

@article{Flambaum_2021,
   title={Radiation from matter-antimatter annihilation in the quark nugget model of dark matter},
   volume={104},
   ISSN={2470-0029},
   url={http://dx.doi.org/10.1103/PhysRevD.104.063042},
   DOI={10.1103/physrevd.104.063042},
   number={6},
   journal={Physical Review D},
   publisher={American Physical Society (APS)},
   author={Flambaum, V. V. and Samsonov, I. B.},
   year={2021},
   month=sep }

@article{PhysRevD.105.123011,
  title = {Thermal and annihilation radiation in the quark nugget model of dark matter},
  author = {Flambaum, V. V. and Samsonov, I. B.},
  journal = {Phys. Rev. D},
  volume = {105},
  issue = {12},
  pages = {123011},
  numpages = {17},
  year = {2022},
  month = {Jun},
  publisher = {American Physical Society},
  doi = {10.1103/PhysRevD.105.123011},
  url = {https://link.aps.org/doi/10.1103/PhysRevD.105.123011}
}

@article{PhysRevD.106.023006,
  title = {Radiation from cold molecular clouds and Sun chromosphere produced by antiquark nugget dark matter},
  author = {Flambaum, V. V. and Samsonov, I. B.},
  journal = {Phys. Rev. D},
  volume = {106},
  issue = {2},
  pages = {023006},
  numpages = {15},
  year = {2022},
  month = {Jul},
  publisher = {American Physical Society},
  doi = {10.1103/PhysRevD.106.023006},
  url = {https://link.aps.org/doi/10.1103/PhysRevD.106.023006}
}

@misc{flambaum2024manifestation,
      title={Manifestation of antiquark nuggets in collisions with the Earth}, 
      author={V. V. Flambaum and I. B. Samsonov and G. K. Vong},
      year={2024},
      eprint={2405.17775},
      archivePrefix={arXiv},
      primaryClass={hep-ph}
}

@misc{expacs_dosemap,
  author       = {{Japan Atomic Energy Agency}},
  title        = {{EXPOSURE DOSE MAP}},
  year         = {2024},
  url          = {https://phits.jaea.go.jp/expacs/dosemap-eng.htm},
  note         = {Accessed: 2024-06-06}
}

@article{magotra2017new,
  title={A new classification scheme of fluorite deposits},
  author={Magotra, Rajni and Namga, Stanzin and Singh, Pawan and Arora, Neha and Srivastava, PK},
  journal={International Journal of Geosciences},
  volume={8},
  number={4},
  pages={599--610},
  year={2017},
  publisher={Scientific Research Publishing}
}

@misc{cirelli2024dark,
      title={Dark Matter}, 
      author={Marco Cirelli and Alessandro Strumia and Jure Zupan},
      year={2024},
      eprint={2406.01705},
      archivePrefix={arXiv},
      primaryClass={hep-ph}
}

@article{PhysRevLett.120.151301,
  title = {Search for Invisible Axion Dark Matter with the Axion Dark Matter Experiment},
  author = {Du, N. and Force, N. and Khatiwada, R. and Lentz, E. and Ottens, R. and Rosenberg, L. J and Rybka, G. and Carosi, G. and Woollett, N. and Bowring, D. and Chou, A. S. and Sonnenschein, A. and Wester, W. and Boutan, C. and Oblath, N. S. and Bradley, R. and Daw, E. J. and Dixit, A. V. and Clarke, J. and O'Kelley, S. R. and Crisosto, N. and Gleason, J. R. and Jois, S. and Sikivie, P. and Stern, I. and Sullivan, N. S. and Tanner, D. B and Hilton, G. C.},
  collaboration = {ADMX Collaboration},
  journal = {Phys. Rev. Lett.},
  volume = {120},
  issue = {15},
  pages = {151301},
  numpages = {5},
  year = {2018},
  month = {Apr},
  publisher = {American Physical Society},
  doi = {10.1103/PhysRevLett.120.151301},
  url = {https://link.aps.org/doi/10.1103/PhysRevLett.120.151301}
}

@misc{vogel2013iaxo,
      title={IAXO - The International Axion Observatory}, 
      author={J. K. Vogel and F. T. Avignone and G. Cantatore and J. M. Carmona and S. Caspi and S. A. Cetin and F. E. Christensen and A. Dael and T. Dafni and M. Davenport and A. V. Derbin and K. Desch and A. Diago and A. Dudarev and C. Eleftheriadis and G. Fanourakis and E. Ferrer-Ribas and J. Galan and J. A. Garcia and J. G. Garza and T. Geralis and B. Gimeno and I. Giomataris and S. Gninenko and H. Gomez and C. J. Hailey and T. Hiramatsu and D. H. H. Hoffmann and F. J. Iguaz and I. G. Irastorza and J. Isern and J. Jaeckel and K. Jakovcic and J. Kaminski and M. Kawasaki and M. Krcmar and C. Krieger and B. Lakic and A. Lindner and A. Liolios and G. Luzon and I. Ortega and T. Papaevangelou and M. J. Pivovaroff and G. Raffelt and J. Redondo and A. Ringwald and S. Russenschuck and J. Ruz and K. Saikawa and I. Savvidis and T. Sekiguchi and I. Shilon and H. Silva and H. H. J. ten Kate and A. Tomas and S. Troitsky and K. van Bibber and P. Vedrine and J. A. Villar and L. Walckiers and W. Wester and S. C. Yildiz and K. Zioutas},
      year={2013},
      eprint={1302.3273},
      archivePrefix={arXiv},
      primaryClass={physics.ins-det}
}

@article{PhysRevLett.112.091302,
  title = {Search for Solar Axions by the CERN Axion Solar Telescope with $^{3}\mathrm{He}$ Buffer Gas: Closing the Hot Dark Matter Gap},
  author = {Arik, M. and Aune, S. and Barth, K. and Belov, A. and Borghi, S. and Br\"auninger, H. and Cantatore, G. and Carmona, J. M. and Cetin, S. A. and Collar, J. I. and Da Riva, E. and Dafni, T. and Davenport, M. and Eleftheriadis, C. and Elias, N. and Fanourakis, G. and Ferrer-Ribas, E. and Friedrich, P. and Gal\'an, J. and Garc\'{\i}a, J. A. and Gardikiotis, A. and Garza, J. G. and Gazis, E. N. and Geralis, T. and Georgiopoulou, E. and Giomataris, I. and Gninenko, S. and G\'omez, H. and G\'omez Marzoa, M. and Gruber, E. and Guth\"orl, T. and Hartmann, R. and Hauf, S. and Haug, F. and Hasinoff, M. D. and Hoffmann, D. H. H. and Iguaz, F. J. and Irastorza, I. G. and Jacoby, J. and Jakov\ifmmode \check{c}\else \v{c}\fi{}i\ifmmode \acute{c}\else \'{c}\fi{}, K. and Karuza, M. and K\"onigsmann, K. and Kotthaus, R. and Kr\ifmmode \check{c}\else \v{c}\fi{}mar, M. and Kuster, M. and Laki\ifmmode \acute{c}\else \'{c}\fi{}, B. and Lang, P. M. and Laurent, J. M. and Liolios, A. and Ljubi\ifmmode \check{c}\else \v{c}\fi{}i\ifmmode \acute{c}\else \'{c}\fi{}, A. and Luz\'on, G. and Neff, S. and Niinikoski, T. and Nordt, A. and Papaevangelou, T. and Pivovaroff, M. J. and Raffelt, G. and Riege, H. and Rodr\'{\i}guez, A. and Rosu, M. and Ruz, J. and Savvidis, I. and Shilon, I. and Silva, P. S. and Solanki, S. K. and Stewart, L. and Tom\'as, A. and Tsagri, M. and van Bibber, K. and Vafeiadis, T. and Villar, J. and Vogel, J. K. and Yildiz, S. C. and Zioutas, K.},
  collaboration = {CAST Collaboration},
  journal = {Phys. Rev. Lett.},
  volume = {112},
  issue = {9},
  pages = {091302},
  numpages = {6},
  year = {2014},
  month = {Mar},
  publisher = {American Physical Society},
  doi = {10.1103/PhysRevLett.112.091302},
  url = {https://link.aps.org/doi/10.1103/PhysRevLett.112.091302}
}

@misc{ortiz2021design,
      title={Design of the ALPS II Optical System}, 
      author={M. Diaz Ortiz and J. Gleason and H. Grote and A. Hallal and M. T. Hartman and H. Hollis and K. S. Isleif and A. James and K. Karan and T. Kozlowski and A. Lindner and G. Messineo and G. Mueller and J. H. Poeld and R. C. G. Smith and A. D. Spector and D. B. Tanner and L. -W. Wei and B. Willke},
      year={2021},
      eprint={2009.14294},
      archivePrefix={arXiv},
      primaryClass={physics.optics}
}

@article{PhysRevLett.118.061302,
  title = {First Results from a Microwave Cavity Axion Search at $24\text{ }\text{ }\ensuremath{\mu}\mathrm{eV}$},
  author = {Brubaker, B. M. and Zhong, L. and Gurevich, Y. V. and Cahn, S. B. and Lamoreaux, S. K. and Simanovskaia, M. and Root, J. R. and Lewis, S. M. and Al Kenany, S. and Backes, K. M. and Urdinaran, I. and Rapidis, N. M. and Shokair, T. M. and van Bibber, K. A. and Palken, D. A. and Malnou, M. and Kindel, W. F. and Anil, M. A. and Lehnert, K. W. and Carosi, G.},
  journal = {Phys. Rev. Lett.},
  volume = {118},
  issue = {6},
  pages = {061302},
  numpages = {5},
  year = {2017},
  month = {Feb},
  publisher = {American Physical Society},
  doi = {10.1103/PhysRevLett.118.061302},
  url = {https://link.aps.org/doi/10.1103/PhysRevLett.118.061302}
}

@inproceedings{lee2020status,
  title={Status of the MADMAX Experiment},
  author={Lee, Chang and MADMAX collaboration},
  booktitle={Microwave Cavities and Detectors for Axion Research: Proceedings of the 3rd International Workshop},
  pages={163--168},
  year={2020},
  organization={Springer}
}

@article{Garcon_2017,
   title={The cosmic axion spin precession experiment (CASPEr): a dark-matter search with nuclear magnetic resonance},
   volume={3},
   ISSN={2058-9565},
   url={http://dx.doi.org/10.1088/2058-9565/aa9861},
   DOI={10.1088/2058-9565/aa9861},
   number={1},
   journal={Quantum Science and Technology},
   publisher={IOP Publishing},
   author={Garcon, Antoine and Aybas, Deniz and Blanchard, John W and Centers, Gary and Figueroa, Nataniel L and Graham, Peter W and Kimball, Derek F Jackson and Rajendran, Surjeet and Sendra, Marina Gil and Sushkov, Alexander O and Trahms, Lutz and Wang, Tao and Wickenbrock, Arne and Wu, Teng and Budker, Dmitry},
   year={2017},
   month=dec, pages={014008} }

@article{Akerib_2013,
   title={The Large Underground Xenon (LUX) experiment},
   volume={704},
   ISSN={0168-9002},
   url={http://dx.doi.org/10.1016/j.nima.2012.11.135},
   DOI={10.1016/j.nima.2012.11.135},
   journal={Nuclear Instruments and Methods in Physics Research Section A: Accelerators, Spectrometers, Detectors and Associated Equipment},
   publisher={Elsevier BV},
   author={Akerib, D.S. and Bai, X. and Bedikian, S. and Bernard, E. and Bernstein, A. and Bolozdynya, A. and Bradley, A. and Byram, D. and Cahn, S.B. and Camp, C. and Carmona-Benitez, M.C. and Carr, D. and Chapman, J.J. and Chiller, A. and Chiller, C. and Clark, K. and Classen, T. and Coffey, T. and Curioni, A. and Dahl, E. and Dazeley, S. and de Viveiros, L. and Dobi, A. and Dragowsky, E. and Druszkiewicz, E. and Edwards, B. and Faham, C.H. and Fiorucci, S. and Gaitskell, R.J. and Gibson, K.R. and Gilchriese, M. and Hall, C. and Hanhardt, M. and Holbrook, B. and Ihm, M. and Jacobsen, R.G. and Kastens, L. and Kazkaz, K. and Knoche, R. and Kyre, S. and Kwong, J. and Lander, R. and Larsen, N.A. and Lee, C. and Leonard, D.S. and Lesko, K.T. and Lindote, A. and Lopes, M.I. and Lyashenko, A. and Malling, D.C. and Mannino, R. and Marquez, Z. and McKinsey, D.N. and Mei, D.-M. and Mock, J. and Moongweluwan, M. and Morii, M. and Nelson, H. and Neves, F. and Nikkel, J.A. and Pangilinan, M. and Parker, P.D. and Pease, E.K. and Pech, K. and Phelps, P. and Rodionov, A. and Roberts, P. and Shei, A. and Shutt, T. and Silva, C. and Skulski, W. and Solovov, V.N. and Sofka, C.J. and Sorensen, P. and Spaans, J. and Stiegler, T. and Stolp, D. and Svoboda, R. and Sweany, M. and Szydagis, M. and Taylor, D. and Thomson, J. and Tripathi, M. and Uvarov, S. and Verbus, J.R. and Walsh, N. and Webb, R. and White, D. and White, J.T. and Whitis, T.J. and Wlasenko, M. and Wolfs, F.L.H. and Woods, M. and Zhang, C.},
   year={2013},
   month=mar, pages={111–126} }

@misc{supercdmscollaboration2022effective,
      title={Effective Field Theory Analysis of CDMSlite Run 2 Data}, 
      author={SuperCDMS Collaboration and M. F. Albakry and I. Alkhatib and D. W. P. Amaral and T. Aralis and T. Aramaki and I. J. Arnquist and I. Ataee Langroudy and E. Azadbakht and S. Banik and C. Bathurst and D. A. Bauer and L. V. S. Bezerra and R. Bhattacharyya and P. L. Brink and R. Bunker and B. Cabrera and R. Calkins and R. A. Cameron and C. Cartaro and D. G. Cerdeño and Y. -Y. Chang and M. Chaudhuri and R. Chen and N. Chott and J. Cooley and H. Coombes and J. Corbett and P. Cushman and F. De Brienne and S. Dharani and M. L. di Vacri and M. D. Diamond and E. Fascione and E. Figueroa-Feliciano and C. W. Fink and K. Fouts and M. Fritts and G. Gerbier and R. Germond and M. Ghaith and S. R. Golwala and J. Hall and N. Hassan and B. A. Hines and M. I. Hollister and Z. Hong and E. W. Hoppe and L. Hsu and M. E. Huber and V. Iyer and A. Jastram and V. K. S. Kashyap and M. H. Kelsey and A. Kubik and N. A. Kurinsky and R. E. Lawrence and M. Lee and A. Li and J. Liu and Y. Liu and B. Loer and P. Lukens and D. B. MacFarlane and R. Mahapatra and V. Mandic and N. Mast and A. J. Mayer and H. Meyer zu Theenhausen and É. Michaud and E. Michielin and N. Mirabolfathi and B. Mohanty and S. Nagorny and J. Nelson and H. Neog and V. Novati and J. L. Orrell and M. D. Osborne and S. M. Oser and W. A. Page and R. Partridge and D. S. Pedreros and R. Podviianiuk and F. Ponce and S. Poudel and A. Pradeep and M. Pyle and W. Rau and E. Reid and R. Ren and T. Reynolds and A. Roberts and A. E. Robinson and H. E. Rogers and T. Saab and B. Sadoulet and I. Saikia and J. Sander and A. Sattari and B. Schmidt and R. W. Schnee and S. Scorza and B. Serfass and S. S. Poudel and D. J. Sincavage and C. Stanford and J. Street and H. Sun and F. K. Thasrawala and D. Toback and R. Underwood and S. Verma and A. N. Villano and B. von Krosigk and S. L. Watkins and O. Wen and Z. Williams and M. J. Wilson and J. Winchell and K. Wykoff and S. Yellin and B. A. Young and T. C. Yu and B. Zatschler and S. Zatschler and A. Zaytsev and E. Zhang and L. Zheng and S. Zuber},
      year={2022},
      eprint={2205.11683},
      archivePrefix={arXiv},
      primaryClass={astro-ph.CO}
}

@article{PLBrink_2009,
doi = {10.1088/1742-6596/150/1/012006},
url = {https://dx.doi.org/10.1088/1742-6596/150/1/012006},
year = {2009},
month = {feb},
publisher = {},
volume = {150},
number = {1},
pages = {012006},
author = {P L Brink and  Z Ahmed and  D S Akerib and  C N Bailey and  D A Bauer and  J Beaty and  R Bunker and  S Burke and  B Cabrera and  D O Caldwell and  J Cooley and  P Cushman and  F DeJongh and  M R Dragowsky and  L Duong and  E Figueroa-Feliciano and  J Filippini and  M Fritts and  S R Golwala and  D R Grant and  J Hall and  R Hennings-Yeomans and  S Hertel and  D Holmgren and  M E Huber and  R Mahapatra and  V Mandic and  K A McCarthy and  N Mirabolfathi and  H Nelson and  L Novak and  R W Ogburn and  M Pyle and  X Qiu and  E Ramberg and  W Rau and  A Reisetter and  T Saab and  B Sadoulet and  J Sander and  R Schmitt and  R W Schnee and  D N Seitz and  B Serfass and  A Sirois and  K M Sundqvist and  A Tomada and  G Wang and  S Yellin and  J Yoo and  B A Young},
title = {The Cryogenic Dark Matter Search (CDMS) experiment: Results and prospects},
journal = {Journal of Physics: Conference Series}
}

@article{Aprile_2023,
   title={First Dark Matter Search with Nuclear Recoils from the XENONnT Experiment},
   volume={131},
   ISSN={1079-7114},
   url={http://dx.doi.org/10.1103/PhysRevLett.131.041003},
   DOI={10.1103/physrevlett.131.041003},
   number={4},
   journal={Physical Review Letters},
   publisher={American Physical Society (APS)},
   author={Aprile, E. and Abe, K. and Agostini, F. and Ahmed Maouloud, S. and Althueser, L. and Andrieu, B. and Angelino, E. and Angevaare, J. R. and Antochi, V. C. and Antón Martin, D. and Arneodo, F. and Baudis, L. and Baxter, A. L. and Bazyk, M. and Bellagamba, L. and Biondi, R. and Bismark, A. and Brookes, E. J. and Brown, A. and Bruenner, S. and Bruno, G. and Budnik, R. and Bui, T. K. and Cai, C. and Cardoso, J. M. R. and Cichon, D. and Cimental Chavez, A. P. and Colijn, A. P. and Conrad, J. and Cuenca-García, J. J. and Cussonneau, J. P. and D’Andrea, V. and Decowski, M. P. and Di Gangi, P. and Di Pede, S. and Diglio, S. and Eitel, K. and Elykov, A. and Farrell, S. and Ferella, A. D. and Ferrari, C. and Fischer, H. and Flierman, M. and Fulgione, W. and Fuselli, C. and Gaemers, P. and Gaior, R. and Gallo Rosso, A. and Galloway, M. and Gao, F. and Glade-Beucke, R. and Grandi, L. and Grigat, J. and Guan, H. and Guida, M. and Hammann, R. and Higuera, A. and Hils, C. and Hoetzsch, L. and Hood, N. F. and Howlett, J. and Iacovacci, M. and Itow, Y. and Jakob, J. and Joerg, F. and Joy, A. and Kato, N. and Kara, M. and Kavrigin, P. and Kazama, S. and Kobayashi, M. and Koltman, G. and Kopec, A. and Kuger, F. and Landsman, H. and Lang, R. F. and Levinson, L. and Li, I. and Li, S. and Liang, S. and Lindemann, S. and Lindner, M. and Liu, K. and Loizeau, J. and Lombardi, F. and Long, J. and Lopes, J. A. M. and Ma, Y. and Macolino, C. and Mahlstedt, J. and Mancuso, A. and Manenti, L. and Marignetti, F. and Marrodán Undagoitia, T. and Martens, K. and Masbou, J. and Masson, D. and Masson, E. and Mastroianni, S. and Messina, M. and Miuchi, K. and Mizukoshi, K. and Molinario, A. and Moriyama, S. and Morå, K. and Mosbacher, Y. and Murra, M. and Müller, J. and Ni, K. and Oberlack, U. and Paetsch, B. and Palacio, J. and Peres, R. and Peters, C. and Pienaar, J. and Pierre, M. and Pizzella, V. and Plante, G. and Qi, J. and Qin, J. and Ramírez García, D. and Singh, R. and Sanchez, L. and dos Santos, J. M. F. and Sarnoff, I. and Sartorelli, G. and Schreiner, J. and Schulte, D. and Schulte, P. and Schulze Eißing, H. and Schumann, M. and Scotto Lavina, L. and Selvi, M. and Semeria, F. and Shagin, P. and Shi, S. and Shockley, E. and Silva, M. and Simgen, H. and Takeda, A. and Tan, P.-L. and Terliuk, A. and Thers, D. and Toschi, F. and Trinchero, G. and Tunnell, C. and Tönnies, F. and Valerius, K. and Volta, G. and Weinheimer, C. and Weiss, M. and Wenz, D. and Wittweg, C. and Wolf, T. and Wu, V. H. S. and Xing, Y. and Xu, D. and Xu, Z. and Yamashita, M. and Yang, L. and Ye, J. and Yuan, L. and Zavattini, G. and Zhong, M. and Zhu, T.},
   year={2023},
   month=jul }

@article{Jorge_Casaus_2009,
doi = {10.1088/1742-6596/171/1/012045},
url = {https://dx.doi.org/10.1088/1742-6596/171/1/012045},
year = {2009},
month = {jun},
publisher = {},
volume = {171},
number = {1},
pages = {012045},
author = {Jorge Casaus},
title = {The AMS-02 experiment on the ISS},
journal = {Journal of Physics: Conference Series}
}

@article{Zhitnitsky_2003,
   title={`Nonbaryonic’ dark matter as baryonic colour superconductor},
   volume={2003},
   ISSN={1475-7516},
   url={http://dx.doi.org/10.1088/1475-7516/2003/10/010},
   DOI={10.1088/1475-7516/2003/10/010},
   number={10},
   journal={Journal of Cosmology and Astroparticle Physics},
   publisher={IOP Publishing},
   author={Zhitnitsky, Ariel R},
   year={2003},
   month=oct, pages={010–010} }

@article{PhysRevD.71.023519,
  title = {Baryon asymmetry, dark matter, and quantum chromodynamics},
  author = {Oaknin, D. H. and Zhitnitsky, A.},
  journal = {Phys. Rev. D},
  volume = {71},
  issue = {2},
  pages = {023519},
  numpages = {19},
  year = {2005},
  month = {Jan},
  publisher = {American Physical Society},
  doi = {10.1103/PhysRevD.71.023519},
  url = {https://link.aps.org/doi/10.1103/PhysRevD.71.023519}
}

@misc{usgs_mrds,
  author = {{U.S. Geological Survey}},
  title = {Mineral Resources Data System (MRDS) - Map by Commodity},
  howpublished = {\url{https://mrdata.usgs.gov/mrds/map-commodity.html}},
  note = {Accessed: 2024-06-11}
}

@article{Forbes:2008uf,
    author = "Forbes, Michael McNeil and Zhitnitsky, Ariel R.",
    title = "{WMAP Haze: Directly Observing Dark Matter?}",
    eprint = "0802.3830",
    archivePrefix = "arXiv",
    primaryClass = "astro-ph",
    reportNumber = "NT@UW-08-05",
    doi = "10.1103/PhysRevD.78.083505",
    journal = "Phys. Rev. D",
    volume = "78",
    pages = "083505",
    year = "2008"
}

@article{amsler1991low,
  title={Low energy antiproton physics},
  author={Amsler, Claude and Myhrer, Fred},
  journal={Annual Review of Nuclear and Particle Science},
  volume={41},
  number={1},
  pages={219--267},
  year={1991},
  publisher={Annual Reviews}
}
\end{document}